\documentclass[a4paper,11pt]{article}
\usepackage{pos}
\usepackage{lineno}
\usepackage[normalem]{ulem}
\usepackage{placeins}

\title{Probing the initial geometry and the onset of collectivity with anisotropic flow in light-ion collisions with ALICE at the LHC}

\author*[\dagger]{Alexian Lejeune}
\onbehalf{ for the ALICE collaboration}

\affiliation[\dagger]{Faculty of Nuclear Sciences and Physical Engineering, Czech Technical University in Prague,\\
  B\v{r}ehová 7, Prague, Czech Republic}

\emailAdd{alexian.lejeune@cern.ch}
\emailAdd{lejeuale@cvut.cz}

\abstract{Collective behavior, traditionally associated with heavy-ion collisions, has been observed in small collision systems at the LHC. The recent oxygen–oxygen and neon–neon collisions allow to bridge the gap between small and large collision systems and to study the collectivity in well-constrained initial states. Measurements of azimuthal anisotropies are sensitive to the initial-state geometry and enable detailed investigations of the nuclear structure of $^{16}$O and $^{20}$Ne, highlighting the interplay between nuclear structure and collective dynamics in ultrarelativistic collisions.

This contribution presents recent anisotropic flow measurements performed by the ALICE Collaboration in OO and Ne–Ne collisions at $\sqrt{s_{\textrm{NN}}}$=5.36 TeV using two- and multiparticle correlation methods. The results are compared to hydrodynamic calculations incorporating the possible nuclear structures of these light nuclei, providing stringent constraints on the modeling of the initial state. In addition, prospects for measuring long-range correlations with the forward ALICE detectors are discussed, with the aim of studying the longitudinal dependence of anisotropic flow and thereby constraining the longitudinal structure of the initial state.}

\FullConference{43rd International Conference on High Energy Physics (ICHEP 2026)\\
30 July  to 5 August , 2026\\
Natal, Brazil\\}

\begin{document}
\maketitle

\section{Introduction}

Anisotropic flow, originally associated with the collective expansion of the medium in heavy-ion collisions, has been observed for more than a decade in smaller collision systems, such as high-multiplicity proton-lead and proton-proton collisions at the LHC \cite{CMS2010,resultsFromFigure}. The origin of collective behavior in small collision systems is still not fully understood. While hydrodynamical models are usually able to reproduce anisotropic flow measurements in heavy-ion collisions, it is a rather challenging task for small collision systems due to the limited understanding of the initial conditions in pp and p--Pb collisions. Other non-fluid dynamical approaches, such as kinetic theory or saturation physics, are also proposed to describe collectivity in smaller collision systems \cite{SmallSystemsDecade}.

In July 2025, a light-ion data-taking campaign was completed at the LHC, during which $^{16}$O and $^{20}$Ne nuclei were collided. Such collisions allow to bridge the gap between small and large collision systems due to their overlapping multiplicity with both small and large collision systems. Moreover, unlike small systems like pp or p--Pb collisions, light ions have a well-defined initial-state geometry. The low-energy nuclear structure models predict that $^{16}$O and $^{20}$Ne exhibit, respectively, tetrahedral and bowlingpin-like configurations, in which nucleons may cluster into $\alpha$ particles \cite{tetrahedralOxygen, nuclearStructureNe20}. Hydrodynamical models implementing the nuclear structure of light ions show that anisotropic flow is sensitive to the deformation of nuclei \cite{QuadrupoleDeformationNeNe}.

\section{Detector setup, dataset and methods}

In this contribution, the anisotropic flow measurements are performed utilizing data from the light-ion runs, OO and Ne--Ne collisions at $\sqrt{s_{\textrm{NN}}} = 5.36$ TeV, recorded with the ALICE detector system \cite{ALICErun3setup}. 

Anisotropic flow coefficients are obtained using different techniques: two-particle correlation functions (2PC) \cite{ALICEIdentifiedParticlePbPb}, a combination of two-particle correlation functions measured with mid- and forward-rapidity detectors (3x2PC) \cite{ALICEpartonicFlow}, or the multiparticle cumulant method \cite{GenericFramework1, GenericFramework2}. Non-flow contributions are suppressed using a low-multiplicity template-fitting method in the 2PC and 3x2PC analyses, and the subevent method in the multiparticle-cumulant analysis.

\section{Results and discussion}

\begin{figure}[htp]
    \centering
    \includegraphics[width=0.40\linewidth]{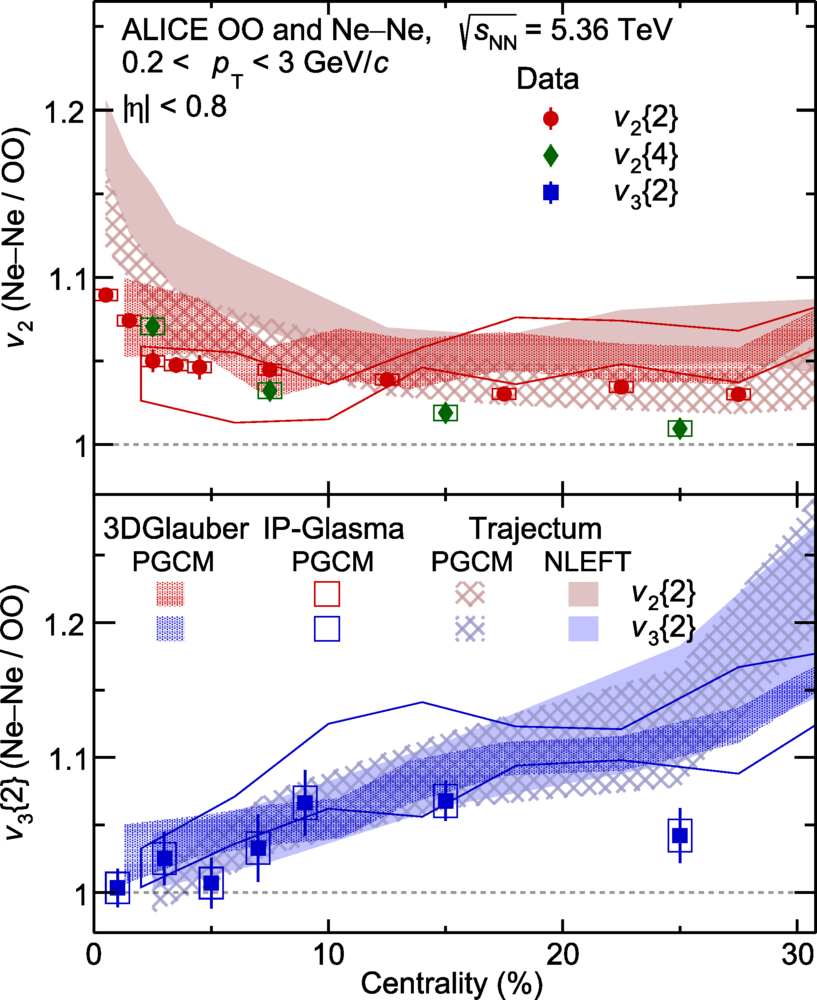}
    \caption{Ratios $v_2\text{(Ne--Ne / OO)}$ and $v_3\text{(Ne--Ne / OO)}$ as a function of centrality in the 0-30\% centrality range  \cite{ALICElightIonsFlow}. The vertical lines represent statistical uncertainties, and the open boxes represent the systematic uncertainties, while most of them are smaller than the symbol size. The measurements are compared with Trajectum calculations with NLEFT and PGCM inputs \cite{nuclearStructureNe20} as well as IP--Glasma + JIMWLK + MUSIC + UrQMD \cite{IPGlasmaAndOtherModels} and 3DGlauber + MUSIC + UrQMD \cite{3DGlauberAndOtherModels} calculations with PGCM input.}
    \label{ratioOONeNe}
\end{figure}

The dependence of the anisotropic flow harmonics ($v_n$) on the centrality has been measured using multiparticle cumulants up to $n=4$ in OO and Ne--Ne collisions  \cite{ALICElightIonsFlow}. To probe the nuclear structures of oxygen and neon, the ratios of elliptic and triangular flow coefficients measured in Ne--Ne and OO collisions are used. This ratio is designed to largely cancel final-state effects arising from the similar system sizes and substantially reduces the uncertainties in both the experimental measurements and model predictions. These ratios are presented in Fig. \ref{ratioOONeNe} and compared with different models implementing the expected nuclear structure of $^{16}$O and $^{20}$Ne \cite{IPGlasmaAndOtherModels,3DGlauberAndOtherModels}. In the most central collisions, the $v_2$ in Ne--Ne collisions is larger than the $v_2$ in OO collisions by around 10\%. For less central collisions, it can be observed that the $v_3$ of OO collisions is higher than that of Ne--Ne collisions. The enhancement in $v_2$ for the most central collision could be attributed to a stronger quadrupole deformation of $^{20}$Ne \cite{QuadrupoleDeformationNeNe} with respect to $^{16}$O. The differences in $v_3$ could originate from the octupole deformation of the oxygen nuclei.

\begin{figure}[htp]
    \centering
    \includegraphics[width=0.85\linewidth]{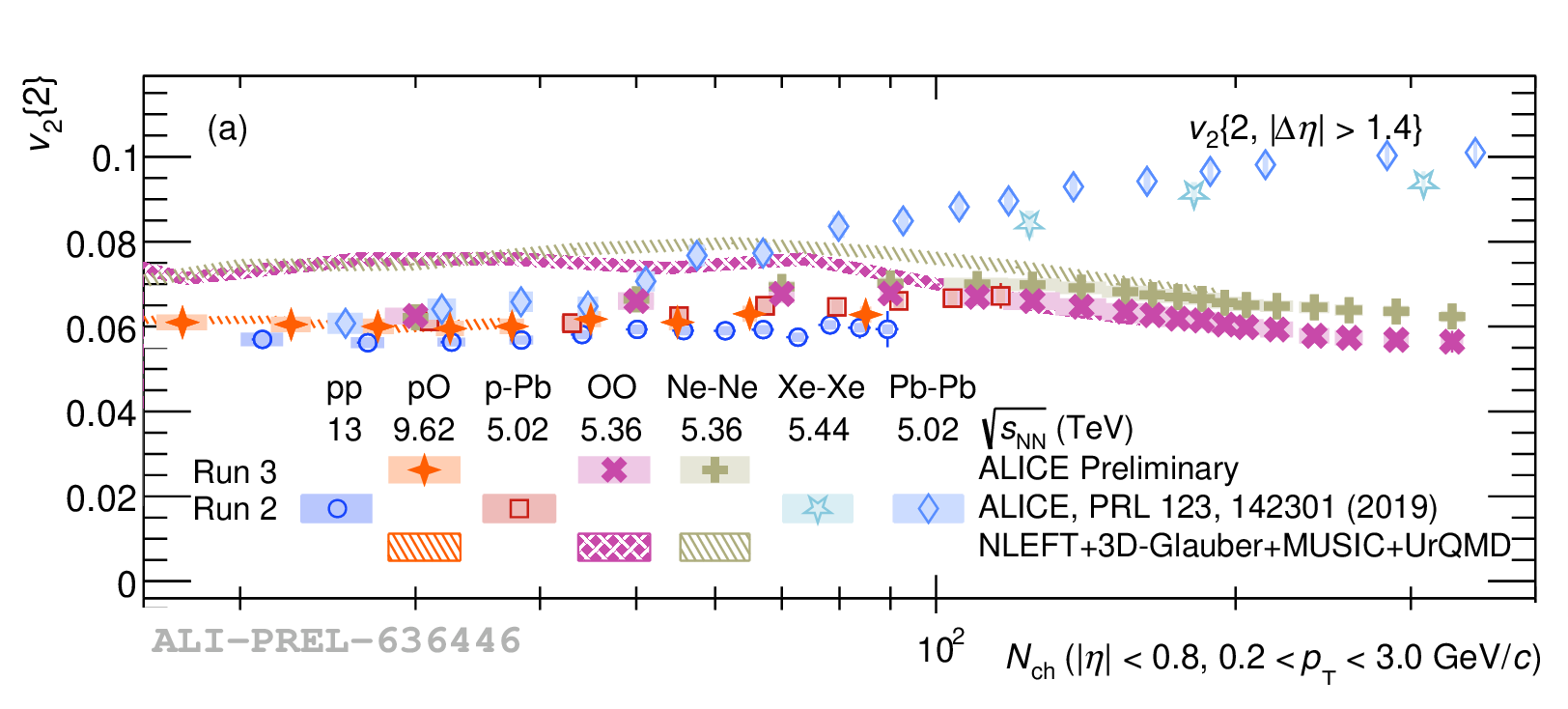}
    \caption{Two-particle elliptic flow coefficients $v_2\{2\}$
 measured as a function of multiplicity in OO and Ne-Ne collisions compared with two-particle anisotropic flow coefficients in Pb--Pb, Xe--Xe, p--Pb and pp  \cite{resultsFromFigure}.}
    \label{multiplicityDependent}
\end{figure}

On top of the nuclear shapes, the flow coefficients can be used to study the onset of collectivity. The overlapping multiplicity ranges of light-ion collisions with smaller and larger collision systems enable comparisons of anisotropic flow harmonics among different collision systems at the LHC, as shown in Fig. \ref{multiplicityDependent}. This figure presents new preliminary results on $v_2$ coefficients in pO, OO and Ne--Ne. For low multiplicity collisions, the amplitude of the elliptic flow is comparable for all collision systems regardless of their size. It is understood that this range is dominated by fluctuations. For collisions with higher multiplicity, $N_{\text{ch}} > 50$, a clear ordering in the amplitude of the $v_2$ coefficient can be observed, with heavy-ion collisions having much larger amplitudes than all the other systems. Geometry effects start to contribute significantly with system size and dominate in Xe--Xe and Pb--Pb collisions. In OO and Ne--Ne collisions, the geometry effect seems to contribute more at higher multiplicity, as $v_2\{2\}$ increases for  $N_{\text{ch}} > 50$, although the increase is much smaller than for Xe--Xe and Pb--Pb collisions. The elliptic flow coefficients decrease after $N_{\text{ch}} > 100$ as the collision geometry becomes less elliptic. Still, a noticeable increase in Ne--Ne with respect to OO collisions can be observed.  When compared with a model implementing NLEFT, 3D-Glauber, MUSIC and UrQMD \cite{3DGlauberAndOtherModels}, the measured elliptic flow of OO and Ne--Ne collisions is well reproduced for $N_{\text{ch}} > 100$, but not at lower multiplicities. For pO collisions, the model describes the data up to $N_{\text{ch}} = 40$.

\begin{figure}[htp]
    \centering
    \includegraphics[width=0.65\linewidth]{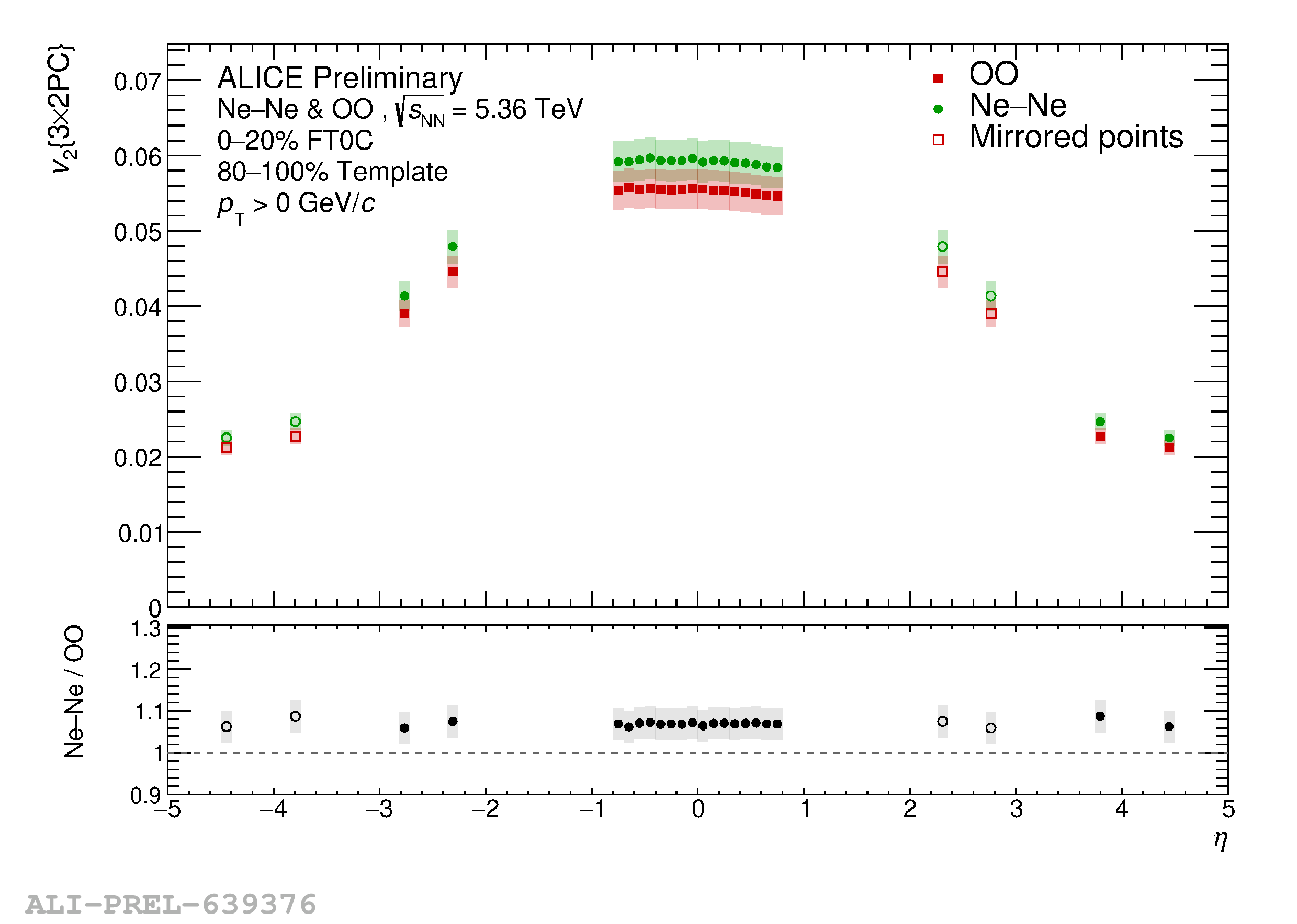}
    \caption{Pseudorapidity dependence of the elliptic flow coefficient $v_2$ in Ne-Ne and OO collisions for centrality 0-20\% using the 3x2PC method. The lower pad displays the ratio of Ne--Ne over OO collisions.}
    \label{longitudinalFlow}
\end{figure}

The longitudinal dependence of the elliptic flow coefficients is crucial to constrain the longitudinal fluctuations of the initial state of the collisions. The pseudorapidity dependence of the elliptic flow coefficient is measured using the mid- and forward-rapidity detectors in ALICE with the 3x2PC method and is shown in Fig. \ref{longitudinalFlow}. It can be observed that the elliptic flow coefficients decrease significatively at large pseudorapidity ranges for light-ion collisions. The enhancement of the $v_2$ coefficient in Ne--Ne collisions with respect to OO collisions is observed up to $\lvert\eta\lvert < 4.9$.


The elliptic flow of identified particles exhibits mass ordering at low $p_{\textrm{T}}$ and baryon-meson grouping and splitting at intermediate $p_{\textrm{T}}$ in heavy-ion collisions. Baryon-meson grouping and splitting at intermediate transverse momentum is considered a characteristic signature of partonic flow in heavy-ion collisions and has also been observed in smaller collision systems \cite{ALICEpartonicFlow}. This feature is understood as an indication that partons individually flow and add up their anisotropic flow as they hadronize through coalescence, explaining why baryons have a higher $v_2$ than mesons. So far, only hydrodynamic models implementing hadronization through coalescence have been able to explain this feature.

\begin{figure}[t]
    \centering
    \includegraphics[width=0.50\linewidth]{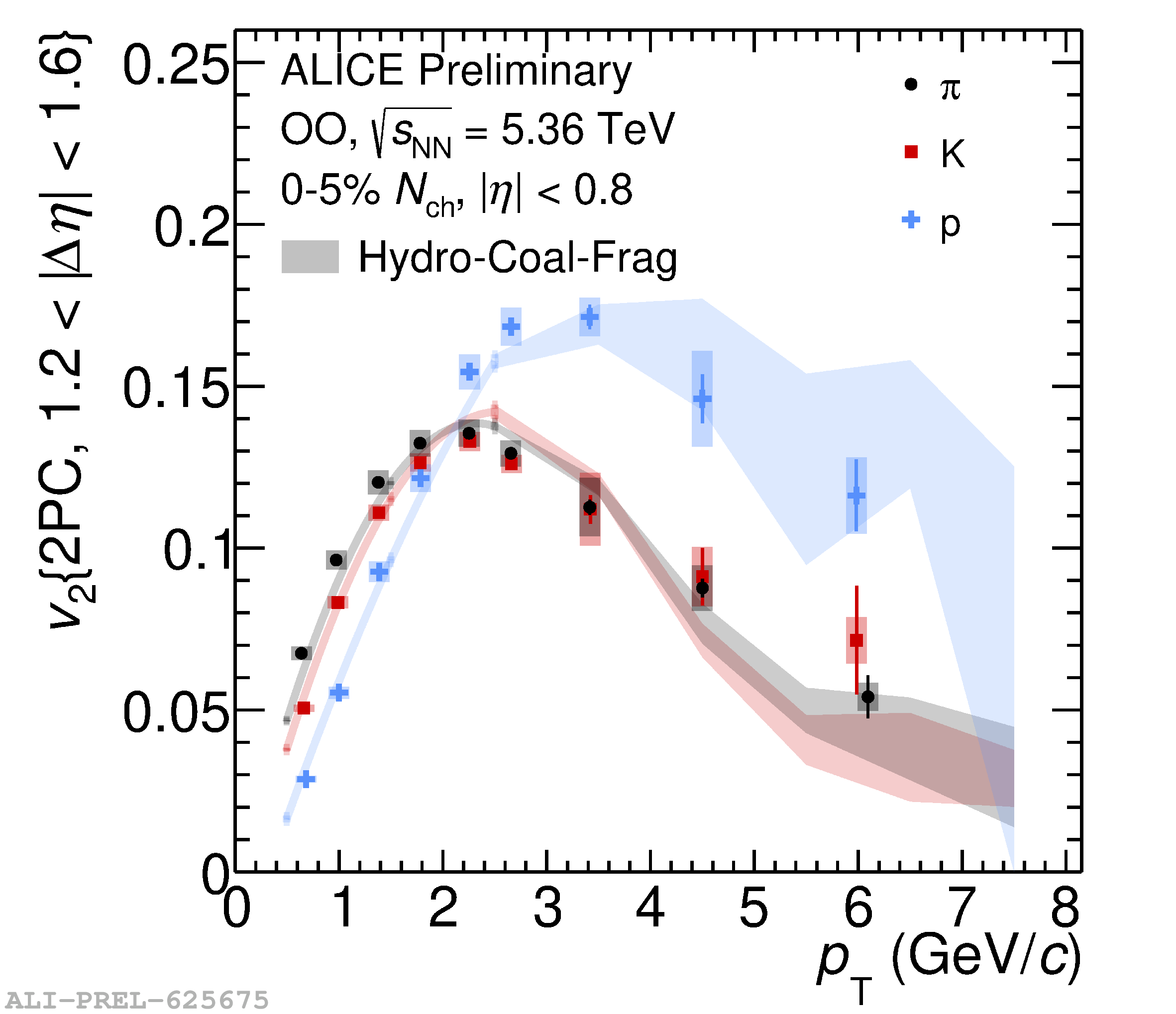}
    \caption{Two-particle elliptic flow coefficients $v_2$ of pions, kaons, and protons in 0-5\% OO collisions compared with the Hydro-Coal-Frag model implementing the nuclear structure of the oxygen.}
    \label{OOpartonicFlowWithHydro}
\end{figure}

Mass ordering and baryon-meson grouping and splitting of light-flavour hadrons are also observed in OO and Ne--Ne collisions. The obtained results in OO collisions are compared to a hydrodynamical model, which implements coalescence in the hadronization stage as well as the nuclear structure of the oxygen nuclei (Fig. \ref{OOpartonicFlowWithHydro}). It can be seen that a reasonable agreement is reached with the data.

\section{Conclusion}

This contribution presented a selection of the latest measurements of anisotropic flow coefficients in light-ion collisions obtained by the ALICE Collaboration. These results shed light on the different nuclear structures of $^{16}$O and $^{20}$Ne nuclei and demonstrated an overall good agreement with hydrodynamical model predictions, specifically in the most central collisions. Moreover, the longitudinal structure of the initial conditions of the collision is investigated through the pseudorapidity dependence of the elliptic flow. The high-multiplicity oxygen-oxygen collisions show characteristic signs of collective partonic flow, similar to both smaller systems (pp, p--Pb) and heavy-ion collision systems. The description of the results by a hydrodynamical model implementing coalescence in the hadronization stage supports the existence of a deconfined medium in light-ion collisions.

\section*{Acknowledgements}

These proceedings were supported by the European Structural and Investment Funds and the Czech Ministry of Education, Youth and Sports (project No. FORTE—CZ.02.01.01/00/22\_008/0004632) and by the Czech Science Foundation under the research project “Study of quark-gluon plasma with hard probes and limits of its formation” no. GA23-07499S.


\bibliographystyle{JHEP}
\bibliography{Annexes/References} 

@article{GenericFramework1,
  title = {Generic framework for anisotropic flow analyses with multiparticle azimuthal correlations},
  author = {Bilandzic, Ante and Christensen, Christian Holm and Gulbrandsen, Kristjan and Hansen, Alexander and Zhou, You},
  journal = {Phys. Rev. C},
  volume = {89},
  issue = {6},
  pages = {064904},
  numpages = {20},
  year = {2014},
  month = {Jun},
  publisher = {American Physical Society},
  doi = {10.1103/PhysRevC.89.064904},
  url = {https://link.aps.org/doi/10.1103/PhysRevC.89.064904}
}

@article{GenericFramework2,
  title = {Generic algorithm for multiparticle cumulants of azimuthal correlations in high energy nucleus collisions},
  author = {Moravcova, Zuzana and Gulbrandsen, Kristjan and Zhou, You},
  journal = {Phys. Rev. C},
  volume = {103},
  issue = {2},
  pages = {024913},
  numpages = {12},
  year = {2021},
  month = {Feb},
  publisher = {American Physical Society},
  doi = {10.1103/PhysRevC.103.024913},
  url = {https://link.aps.org/doi/10.1103/PhysRevC.103.024913}
}

@ARTICLE{ALICEpartonicFlow,
  title     = "Observation of partonic flow in proton-proton and proton-nucleus
               collisions",
  author    = "{ALICE Collaboration}",
  collaboration = {ALICE}, 
    doi = {10.1038/s41467-025-67795-1},
    url = {https://doi.org/10.1038/s41467-025-67795-1},
  journal   = "Nat. Commun.",
  publisher = "Springer Science and Business Media LLC",
  volume    =  17,
  number    =  1,
  month     =  mar,
  year      =  2026,
  copyright = "https://creativecommons.org/licenses/by/4.0",
  language  = "en"
}

@article{tetrahedralOxygen,
  title = {Evidence for the Tetrahedral Nature of $^{16}\mathrm{O}$},
  author = {Robson, D.},
  journal = {Phys. Rev. Lett.},
  volume = {42},
  issue = {14},
  pages = {876--879},
  numpages = {0},
  year = {1979},
  month = {Apr},
  publisher = {American Physical Society},
  doi = {10.1103/PhysRevLett.42.876},
  url = {https://link.aps.org/doi/10.1103/PhysRevLett.42.876}
}

@article{nuclearStructureNe20,
title = {Influence of left-right asymmetry degrees of freedom in self-consistent calculations of 20Ne},
journal = {Nuclear Physics A},
volume = {410},
number = {1},
pages = {125-136},
year = {1983},
issn = {0375-9474},
doi = {https://doi.org/10.1016/0375-9474(83)90405-0},
url = {https://www.sciencedirect.com/science/article/pii/0375947483904050},
author = {S. Marcos and H. Flocard and P.H. Heenen}
}

@article{QuadrupoleDeformationNeNe,
  title = {Exploiting $^{20}\mathrm{Ne}$ Isotopes for Precision Characterizations of Collectivity in Small Systems},
  author = {Giacalone, Giuliano and Bally, Benjamin and Nijs, Govert and Shen, Shihang and Duguet, Thomas and Ebran, Jean-Paul and Elhatisari, Serdar and Frosini, Mikael and L{\"a}hde, Timo A. and Lee, Dean and Lu, Bing-Nan and Ma, Yuan-Zhuo and Mei{\ss{}}ner, Ulf-G. and Noronha-Hostler, Jacquelyn and Plumberg, Christopher and Rodr{\'{\i}}guez, Tom{\'a}s R. and Roth, Robert and van der Schee, Wilke and Som{\`a}, Vittorio},
  journal = {Phys. Rev. Lett.},
  volume = {135},
  issue = {1},
  pages = {012302},
  numpages = {9},
  year = {2025},
  month = {Jul},
  publisher = {American Physical Society},
  doi = {10.1103/k8rb-jgvq},
  url = {https://link.aps.org/doi/10.1103/k8rb-jgvq}
}

@article{IPGlasmaAndOtherModels,
  title = {Collision-Energy Dependence in Heavy-Ion Collisions from Nonlinear QCD Evolution},
  author = {M{\"a}ntysaari, Heikki and Schenke, Bj{\"o}rn and Shen, Chun and Zhao, Wenbin},
  journal = {Phys. Rev. Lett.},
  volume = {135},
  issue = {2},
  pages = {022302},
  numpages = {8},
  year = {2025},
  month = {Jul},
  publisher = {American Physical Society},
  doi = {10.1103/gf4y-p5j7},
  url = {https://link.aps.org/doi/10.1103/gf4y-p5j7}
}

@article{3DGlauberAndOtherModels,
  title = {Bayesian analysis of $(3+1)\mathrm{D}$ relativistic nuclear dynamics with the RHIC beam energy scan data},
  author = {Jahan, Syed Afrid and Roch, Hendrik and Shen, Chun},
  journal = {Phys. Rev. C},
  volume = {110},
  issue = {5},
  pages = {054905},
  numpages = {17},
  year = {2024},
  month = {Nov},
  publisher = {American Physical Society},
  doi = {10.1103/PhysRevC.110.054905},
  url = {https://link.aps.org/doi/10.1103/PhysRevC.110.054905}
}

@misc{SmallSystemsDecade,
      title={A Decade of Collectivity in Small Systems}, 
      author={Jan Fiete Grosse-Oetringhaus and Urs Achim Wiedemann},
      year={2024},
      doi = {2407.07484},
      eprint={2407.07484},
      archivePrefix={arXiv},
      primaryClass={hep-ex},
      url={https://arxiv.org/abs/2407.07484}, 
}

@article{CMS2010,
  title     = "Observation of long-range, near-side angular correlations in
               proton-proton collisions at the {LHC}",
  author    = "",
  
  collaboration = {CMS},
  journal   = "J. High Energy Phys.",
  publisher = "Springer Science and Business Media LLC",
  volume    =  2010,
  number    =  9,
  month     =  sep,
  year      =  2010,
  doi       = "https://doi.org/10.1007/JHEP09(2010)091",
  language  = "en"
}

\end{document}